\documentclass[12pt]{iopart}

\usepackage[]{graphicx}

\begin{document}

\title[]{Multiphoton transitions for delay-zero calibration in attosecond spectroscopy}

\author{J Herrmann$^1$,  M Lucchini$^1$, S Chen$^2$, M Wu$^2$, A Ludwig$^1$, \mbox{L Kasmi$^1$}, K J Schafer$^2$, L Gallmann$^{1,3}$, M B Gaarde$^2$ and \mbox{U Keller$^1$}}

\address{$^1$Department of Physics, Institute of Quantum Electronics, ETH Zurich, CH-8093 Z\"urich,
Switzerland}
\address{$^2$Department of Physics and Astronomy, Louisiana State University, Baton Rouge, Louisiana 70803, USA}
\address{$^3$Institute of Applied Physics, University of Bern, CH-3012 Bern, Switzerland}

\ead{jens.herrmann@phys.ethz.ch}

\begin{abstract}
The exact delay-zero calibration in an attosecond pump-probe experiment is important for the correct interpretation of experimental data. In attosecond transient absorption spectroscopy the determination of the delay-zero exclusively from the experimental results is not straightforward and may introduce significant errors. Here, we report the observation of quarter-laser-cycle (4$\omega$) oscillations in a transient absorption experiment in helium using an attosecond pulse train overlapped with a precisely synchronized, moderately strong infrared pulse. We demonstrate how to extract and calibrate the delay-zero with the help of the highly nonlinear 4$\omega$ signal. A comparison with the solution of the time-dependent Schr\"odinger equation is used to confirm the accuracy and validity of the approach. Moreover, we study the mechanisms behind the quarter-laser-cycle and the better-known half-laser-cycle oscillations as a function of experimental parameters. This investigation yields an indication of the robustness of our delay-zero calibration approach. 
\end{abstract}

\vspace{2pc}

\maketitle

\section{Introduction}
Transient absorption spectroscopy plays a major role in the fast evolving field of attosecond science. It allows observing electron dynamics of atoms, molecules and solids on their natural timescale, with an all-optical method \cite{Gallmann13}. For the first demonstrations of attosecond transient absorption spectroscopy the transmitted XUV radiation was detected after a noble gas target using either a single attosecond pulse (SAP) or an attosecond pulse train (APT) in the extreme ultraviolet (XUV) spectral range which was overlapped with a femtosecond infrared (IR) pulse \cite{Goulielmakis10,Wang10,Holler11}. Hence, this technique benefits from the numerous advantages of photon detection over the detection of charged particles as discussed in more details in \cite{Gallmann13}. In addition to the fast data acquisition due to charge-coupled device (CCD) based spectrometers and the absence of space charge effects, it is possible to examine bound-bound transitions which stay hidden in conventional charged-particle detection experiments. 

A key issue for the interpretation of the experimental data is the correct determination of the delay-zero, where the maximum of the envelope of the IR pulse and the SAP or APT exactly overlap. The precision of delay-zero calibration has to match the timescales of the dynamics being studied and the time resolution of the experiment. As we will show below, just "reading" and "interpreting" experimental data in a straightforward way as done successfully before for femtosecond transient absorption spectroscopy contains serious pitfalls and may lead to errors in finding delay-zero on the order of several femtoseconds. Without theoretical support or other knowledge about the processes being studied, delay-zero can in general not be extracted from experimental data with the required precision. A variety of recent publications have discussed the absorption of XUV radiation in helium (He) around its first ionization potential, either using SAPs or APTs \cite{Chini12,Chen12,Chini13,Pfeiffer13,Lucchini13,Herrmann13}. Due to the diversity of the observed effects in these experiments, \textit{e.g.} sub-cycle AC Stark shift, light-induced states etc., it is not obvious if any of these effects can be used to define delay-zero. In this paper, we combine experimental results with calculations to identify a nonlinear light-matter interaction that supplies the proper delay-zero. 

Our study yields a purely experimental method for calibrating the delay-zero in an attosecond transient absorption experiment using an APT synthesized from a number of high-order harmonics (HHs) of an infrared laser field. We will show that neither the maximum of the total absorption, nor the envelope of the already well-known and discussed half-laser-cycle (2$\omega$) oscillations are suitable for this purpose \cite{Holler11,Chini12,Chen12,Chini13,Lucchini13,Chen12_2,Chen13,Wang13}. Here and throughout this work, $\omega$ represents the frequency of the IR field. On the other hand, we report the first experimental observation of quarter-laser-cycle (4$\omega$) oscillations in the transmitted XUV radiation as a function of the delay between an APT and an IR pulse and show that the maximum of the 4$\omega$-oscillations coincides with the delay-zero. In all presented figures, we use the maximum of the 4$\omega$-oscillations in the absorption of the 13\textsuperscript{th} harmonic (HH 13) to define the delay-zero. We discuss the parameters needed for the manifestation of the 4$\omega$-oscillations and demonstrate that this highly nonlinear effect enables us to accurately define delay-zero. Moreover, we systematically study the influence of the IR intensity on the 2$\omega$- and 4$\omega$-oscillations. This systematic study defines the robustness of our proposed method.

In section 2, we begin with a short overview, experimental details, a general discussion on 2$\omega$- and 4$\omega$-oscillations and introduce the theoretical model. In section 3, we describe how we use the 4$\omega$-oscillations to experimentally calibrate the delay-zero and show that our experimental results are in excellent agreement with the theoretical predictions. Section 4 presents a systematic investigation of the dependence of 2$\omega$- and 4$\omega$-oscillations on IR intensity. Finally, we compare our transient absorption results with a measurement of the He\textsuperscript{+} ion yield in section 5, and conclude with section 6.

\section{Quarter-laser-cycle oscillations}

In 2007, Johnsson and co-workers published an experimental and theoretical pump-probe study using an APT combined with a moderately strong IR pulse in He \cite{Johnsson07}. "Moderately strong" corresponds to an intensity of approximately 10$^{13}$\,W/cm$^2$ which is insufficient to excite electrons out of the ground state, but substantial enough to deform the atomic potential. They investigated the He\textsuperscript{+} ion yield as a function of the APT-IR delay and discovered 2$\omega$-oscillations of the total ion yield. Their results triggered several detailed studies of the same 2$\omega$-oscillations by means of attosecond transient absorption spectroscopy \cite{Holler11,Chini12,Chen12,Chini13,Lucchini13,Chen12_2,Chen13,Wang13}. The mechanism giving rise to the 2$\omega$-oscillations involves the so-called 'transient virtual states' initiated by the IR field. These originate from two-color absorption processes with one XUV photon and a variable number of IR photons \cite{Chen12,Chini13}. These different excitation pathways interfere destructively or constructively, depending on the APT-IR delay. As a result, the absorption probability exhibits 2$\omega$-oscillations. Additionally, 2$\omega$-oscillations can also originate from the two-photon coupling between real states \cite{Chen12_2}. 

\begin{figure}[t]
\begin{center}
 \begin{tabular}{c}
\includegraphics[width=14cm]{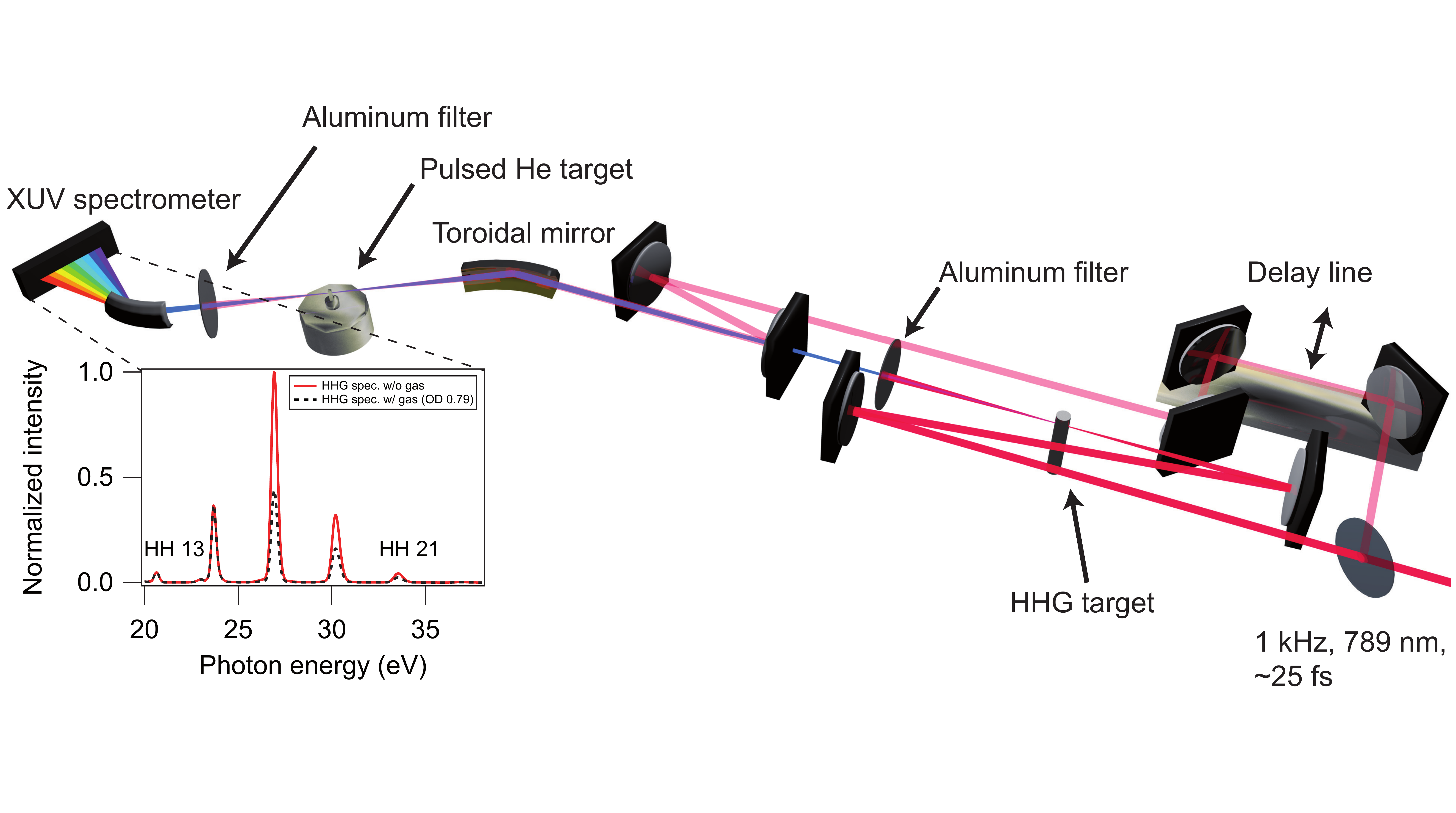}
 \end{tabular}
\end{center}
\caption[example]{\label{fig:Figure_setup} Schematic of the experimental setup \cite{Locher14}. Infrared (IR) pulses with duration of 25\,fs and central wavelength of 789\,nm are used to generate an an attosecond pulse train (APT) via high harmonic generation (HHG) in a xenon gas target. A small fraction of the fundamental IR beam is split off before the HHG and sent over a delay line. After the HHG the residual IR radiation is blocked with a 100-nm thick aluminum filter. The APT and IR beams are recombined with the help of a mirror with a center hole and focused by a toroidal mirror into the pulsed He target. After the interaction target, the IR beam is blocked with a second aluminum filter and the transmitted radiation is detected with an XUV spectrometer. The inset shows the spectral shape of the APT without He gas in the target (red-solid) and with He gas (black-dashed) with an optical density of 0.79 at 26.7\,eV. }
\end{figure}

Theoretical work by Chen \textit{et al.} \cite{Chen12_2} discussed the transient absorption of an APT in laser-dressed He atoms and predicted the occurrence of oscillations with a new periodicity, which appears in neither the initial APT nor the IR field, namely 4$\omega$-oscillations. These oscillations originate from multiphoton coupling of HHs constituting the APT that are spaced four IR photons apart, \textit{e.g.}, HH 13 connects to HH 17. This means that energy between these HHs is exchanged via a four-photon process. Moreover, the authors discussed the influence of resonances on the nonlinear coupling. The calculations reveal that a HH being in resonance with an excited state increases its nonlinear coupling to other HHs. As a result the coupling strength of the HHs depends on the driving IR wavelength and intensity. The 4$\omega$-oscillations were recently observed for the first time in an experiment using a SAP instead of an APT, but they were not discussed any further \cite{Chini13}.  

Here, we investigate the 4$\omega$-oscillations in an attosecond transient absorption experiment with APTs. Figure 1 shows the experimental setup. A more detailed description can be found in \cite{Locher14}. We use the main part of the output from a Ti:sapphire-based laser amplifier system to generate the APT via high-order harmonic generation (HHG). For this purpose, we focus the beam into a 3-mm long gas cell filled with xenon (Xe). After the generation of the APT we block the co-propagating residual IR radiation with a 100-nm thick aluminum filter. Additionally, the aluminum filter compresses the pulses in the APT in time \cite{Lopez05}. In front of the HHG target we separate off a small fraction of the fundamental IR beam and send it along an independent beam path that includes an optical delay line for varying the delay between the APT and the IR. The two beams are then recombined using a mirror with a hole in the center. This generation scheme results in an inherent synchronization of the APT to the IR pulses. After recombination, a toroidal mirror focuses both beams into the interaction gas target. A motorized iris in the IR arm is used to adjust the IR intensity in the interaction region. The pulsed gas target operates at the laser repetition rate of 1\,kHz with an opening time of ~60\,$\mu$s per laser shot. The optical density is adjusted through the backing pressure of the pulsed target. In this work we use an optical density of 0.79 for HH 17 at 26.7\,eV. After the interaction, another metallic filter removes the IR radiation and we detect the transmitted spectrum with an XUV spectrometer. The spectrometer provides a resolution of $\approx$50\,meV in the region of interest. The inset of figure 1 depicts the spectrum of the APT with and without gas but no IR present. While HH 17 and higher are located above the ionization potential (24.59\,eV) and strongly absorbed in the presence of He, HH 13 and 15 are essentially transmitted unchanged \cite{Wiese09}. 

\begin{figure}[t]
\begin{center}
 \begin{tabular}{c}
\includegraphics[width=14cm]{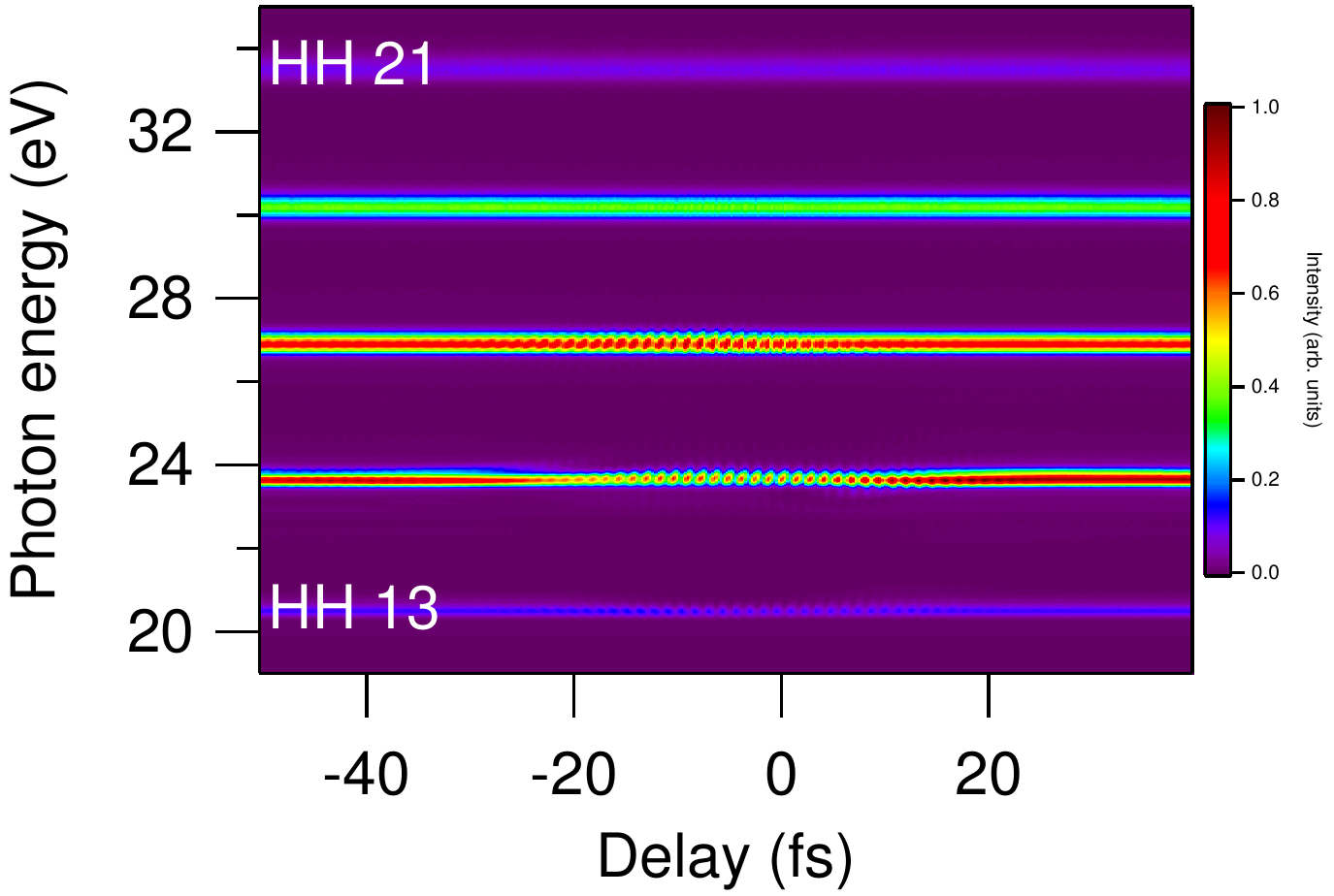}
 \end{tabular}
\end{center}
\caption[example]{\label{fig:Figure_delayscan} Transmitted XUV intensity as a function of photon energy and APT-IR delay. The APT consists of the HHs 13 to 21. The delay scan was recorded with IR intensity of $2.6\cdot10^{12}$\,W/cm$^2$ and at an optical density of 0.79 at 26.7\,eV in the He target. For negative delays the APT is preceding the IR pulses. The 2$\omega$-oscillations in the HH are visible to the naked eye.}
\end{figure}

Figure 2 depicts a delay scan showing the transmitted XUV spectral power density color-coded against the photon energy and the APT-IR delay. The APT primarily consists of HHs 13 to 21. The delay scan was recorded with an IR intensity of $2.6\cdot10^{12}$\,W/cm$^2$. The delay step size is 0.2\,fs. The 2$\omega$-oscillations, especially in HHs 13 to 17, are clearly visible to the naked eye. The delay-zero here and in all the following figures was calibrated with the temporal envelope of the 4$\omega$-oscillations of HH 13, as  discussed in section 3.

To study the oscillations in the transmitted XUV radiation in more detail we have to be able to quantify the oscillation strength of different delay scans. Thus, we define in a first step the energy-integrated absorption $\Pi(\tau)$ as a function of the APT-IR delay $\tau$: 

\begin{equation}
\label{eq:eia}
\Pi(\tau)=-\frac{\int_{\delta E}(T(E,\tau)-T_0(E))\rmd E}{\int_{\delta E'}(S(E)-T_0(E))\rmd E}
\end{equation}

where $E$ is the photon energy, $S(E)$ is the HHG spectrum before the interaction with the gas target, and $T_0(E)$ and $T(E,\tau)$ represent the transmitted XUV radiation when the IR field is turned off and on at a fixed delay $\tau$, respectively. The energy integration in the numerator over the spectral window $\delta E$ is performed around each HH. For the normalization, we integrate in the denominator over the three HHs, 13 to 17. HH 19 and 21 are not very sensitive to the IR field and therefore not included in the normalization. A positive value of $\Pi(\tau)$ corresponds to absorption induced by the IR field, while a negative value means a net emission of photons. As an example we show in figure 3 the energy-integrated absorption of HH 13 (energy integration window $\Delta E=0.72$\,eV), 15 ($\Delta E=1.17$\,eV) and 17 ($\Delta E=1.08$\,eV) for the delay scan shown in figure 2. In this representation of the experimental data the 2$\omega$-oscillations become even more apparent. HH 13 and 15 exhibit an increase of absorption around delay-zero. This increase corresponds to multiphoton absorption of one XUV and one or more IR photons. HH 17 shows only oscillations and no IR-induced multiphoton absorption due to being energetically located above the first ionization potential of He. HH 17 shows, however, a first hint of 4$\omega$-oscillations around delay-zero.

\begin{figure}[t]
\begin{center}
 \begin{tabular}{c}
\includegraphics[width=14cm]{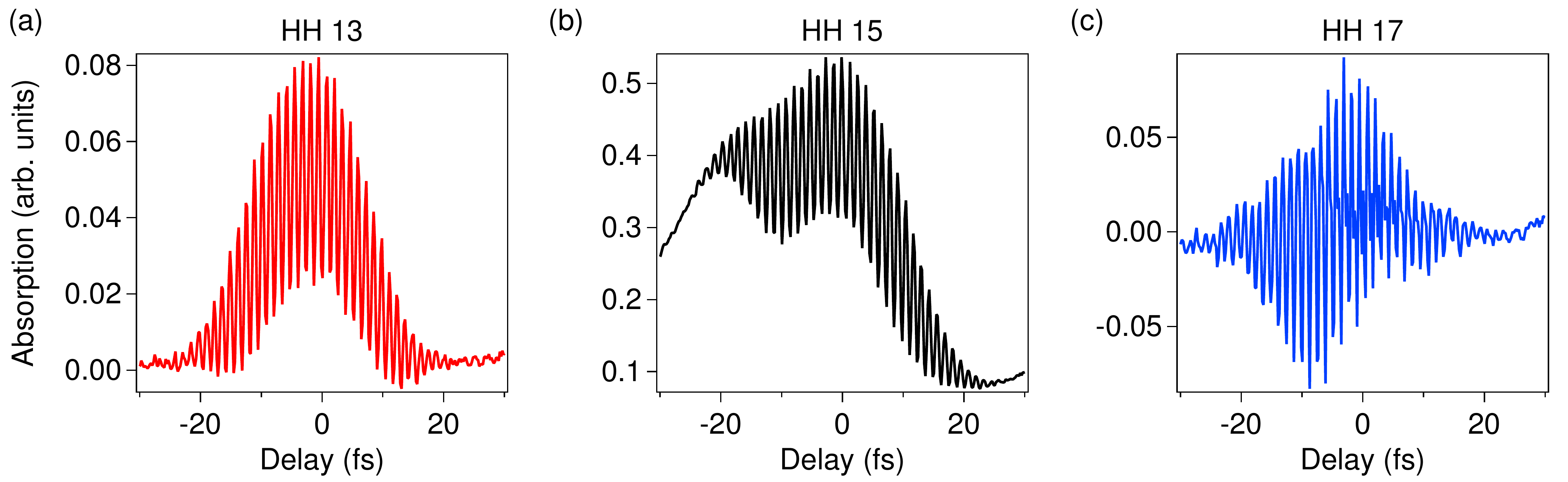}
 \end{tabular}
\end{center}
\caption[example]{\label{fig:Figure_eeia} (a) Energy-integrated absorption $\Pi(\tau)$ of HH 13, (b) 15 and (c) 17 recorded at a laser intensity of $2.6\cdot10^{12}$\,W/cm$^2$ and an optical density of 0.79 at 26.7\,eV. For HH 13 and 15 the absorption is enhanced when XUV and IR radiation overlap due to multiphoton ionization. HH 17 exhibits only oscillations and no global enhancement of the absorption. The 4$\omega$-oscillations are already weakly visible in HH 17 around delay-zero.}
\end{figure}

In a second step, we disentangle the different multiphoton contribution in the one-dimensional energy-integrated absorption $\Pi(\tau)$ by applying a Gaussian-Wigner time-frequency transform that yields a two-dimensional function of frequency and APT-IR delay. The Gaussian-Wigner transform uses the Wigner transform of $\Pi(\tau)$ \cite{Wigner32}, which reads as:

\begin{equation}
W(t,\nu)=\int_{-\infty}^{+\infty}U(t+x'/2)U^*(t-x'/2)\rme^{-2\pi\rmi x'\nu}\rmd x'
\end{equation}

and convolves it with a two-dimensional Gaussian window:

\begin{equation}
GWT(t,\nu;\delta_t,\delta_\nu)=\frac{1}{\delta_\nu\cdot\delta_t}\int\int\rmd  t'\rmd \nu'W(t',\nu')\rme^{-2\pi(\frac{t-t'}{\delta_t})^2}\rme^{-2\pi(\frac{\nu-\nu'}{\delta_\nu})^2}
\end{equation}

For the window in the time domain we use $\delta_{\tau}=5$\,fs ($\approx$1.9 optical cycles of the IR field) and the frequency window is defined by $\delta_{\nu}\cdot\delta_{\tau}=1$, which leads to $\delta_{\nu}=0.2$\,PHz. This representation allows us to extract which frequency components appear at which delay. Figures 4(a)-(c) show the Gaussian-Wigner transform of HHs 13 to 17 corresponding to the traces depicted in figure 3. Besides the 2$\omega$-oscillations at a frequency of 0.76\,PHz we now also observe 4$\omega$-oscillations at 1.52\,PHz. The period of the 4$\omega$-oscillation at a driving wavelength of 789\,nm is 660\,as. In HH 15 the 4$\omega$-oscillations are relatively weak compared to the 2$\omega$-oscillations. This is because HH 15 does not couple to HH 11, which is not part of our APT spectrum. Therefore, we will restrict our discussion of the 4$\omega$-oscillations in the absorption of HH 13 and HH 17.

\begin{figure}[t]
\begin{center}
 \begin{tabular}{c}
\includegraphics[width=14cm]{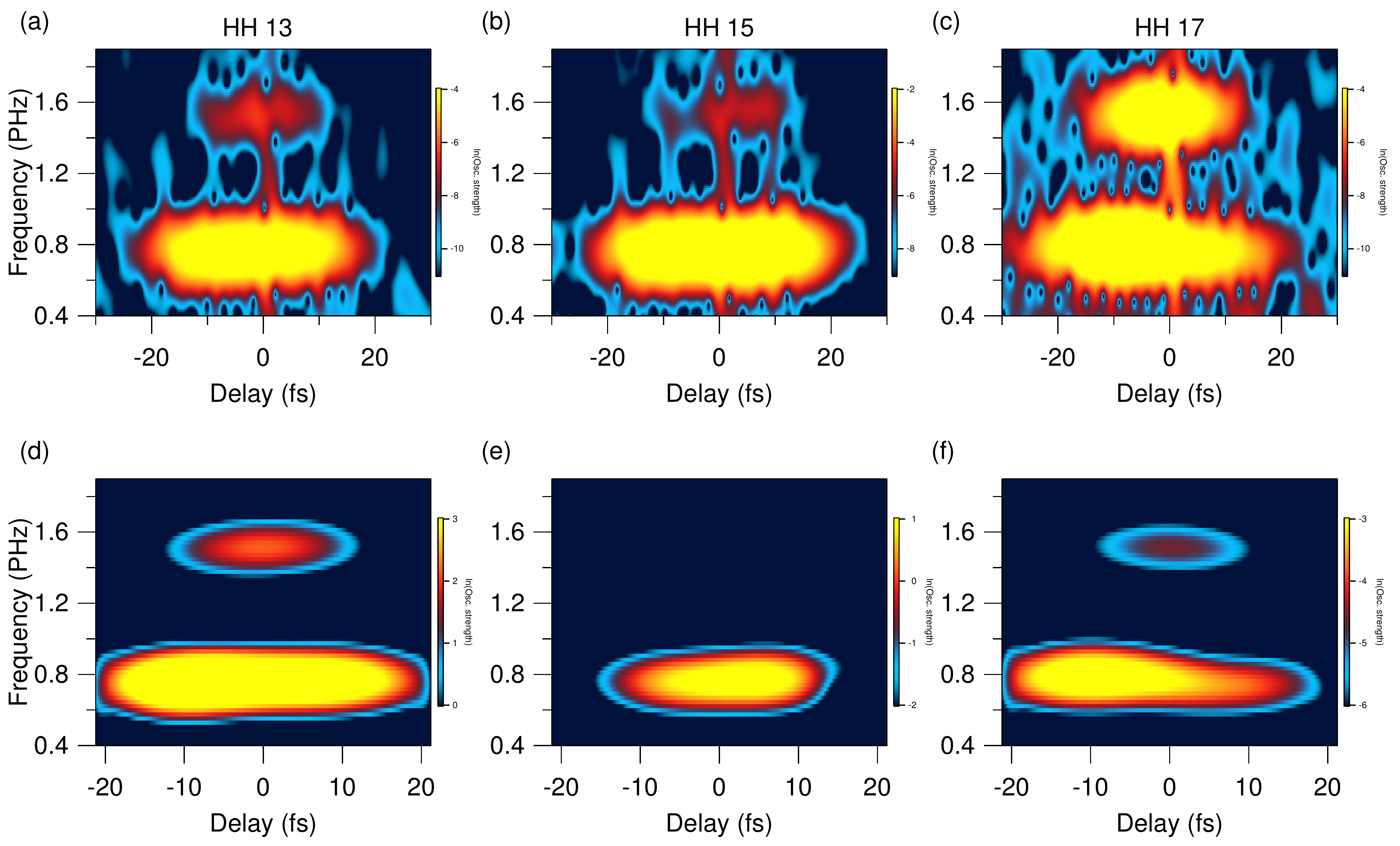}
 \end{tabular}
\end{center}
\caption[example]{\label{fig:Figure_GaussWigner} (a)-(c) Gaussian-Wigner transform of the energy-integrated absorption of HH 13, 15 and 17 taken from figures 3(a)-(c), respectively. The signal at 0.76\,PHz corresponds to 2$\omega$-oscillations, which are also visible in figures 2 and 3. Additionally, we observe a signal at 1.52\,PHz, which corresponds to 4$\omega$-oscillations.  (d)-(f) present the corresponding calculated Gaussian-Wigner transform of the macroscopic absorption probabilities at an IR intensity of $6.0\cdot10^{12}$\,W/cm$^2$. The density of the He atoms is $3.3\cdot10^{17}$\,cm$^{-3}$. The reader should note the different scaling of the delay axis for the experimental and calculated results. }
\end{figure}

We have also investigated theoretically the ultrafast transient absorption of the APT in He. We have calculated both the microscopic (single atom) and macroscopic absorption probabilities for each of the harmonics in the APT as a function of the APT-IR delay, as described in detail in [19]. Briefly, at the single atom level we compute the response function $S(\omega)=2\Im[d(\omega)E^*(\omega)]$, where $d(\omega)$ and $E(\omega)$ are the Fourier transforms of the time-dependent dipole moment and the full APT-IR electric field, respectively. The time-dependent dipole moment is calculated by direct numerical integration of the time-dependent Schr\"odinger equation (TDSE) in the single active electron approximation \cite{Gaarde11}. $S(\omega)$ is the absorption probability per frequency, so that the probability to absorb a certain harmonic is the integral of $S(\omega)$ over the bandwidth of that harmonic. For the macroscopic calculations we numerically solve the coupled TDSE and MaxwellÕs wave equation \cite{Gaarde11} which yields the space- and time-dependent electric field at the end of the He gas jet. From this we calculate the same energy integrated absorption probability as in (\ref{eq:eia}). In all of the calculations we use an APT synthesized from harmonics 13 through 21, with initial relative strengths of 0.25, 0.6, 1, 0.6 and 0.25, and all initially in phase. The full width at half maximum (FWHM) duration of the APT is 11\,fs, and the peak intensity is $7.0\cdot10^{10}$\,W/cm$^2$. The IR pulse has a central wavelength of 795\,nm, a FWHM pulse duration of 25\,fs, and a peak intensity which varies between $1.0\cdot10^{12}$\,W/cm$^2$ and $10.0\cdot10^{12}$\,W/cm$^2$.  

Figures 4(d)-(f) show the Gaussian-Wigner transform of the calculated macroscopic absorption probabilities for harmonics 13, 15, and 17, for an IR intensity of $6.0\cdot10^{12}$\,W/cm$^2$ and He density of $3.3\cdot10^{17}$\,cm$^{-3}$. The theory agrees well with the experiment, with the 2$\omega$-oscillations being asymmetric around delay-zero (especially in HH 17) whereas the 4$\omega$-oscillations of HH 13 and HH 17 exhibit a very stable maximum at delay-zero. The 4$\omega$-oscillations of HH 15 are very weak, also in agreement with the experimental result. We note that at this density, the macroscopic absorption probabilities are very similar to the single atom absorption probabilities. We remark that the difference in strength between the 4$\omega$-oscillations of HH 13 and HH 17 is due to the normalization in (\ref{eq:eia}). In the calculations, the raw strengths of the 4$\omega$-oscillations of HH 13 and HH 17 match exactly since the only four-IR-photon coupling of HH 13 is to HH 17 and vice versa (given that the absorption of HH 21 is very weak). 

\section{Delay-zero calibration}

The simplest method for the definition of the APT-IR delay-zero is to use the maximum of the energy-integrated total absorption. Figure 5(a) shows the energy-integrated total absorption of the APT integrated in energy from 19.5\,eV to 38.5\,eV for an IR intensity of $1.3\cdot10^{12}$\,W/cm$^2$. For the fit function, we choose the sum of a Gaussian and a linear function to take into account that the total absorption does not have the same value for large negative and positive delays. For large positive delays the preceding IR pulse does not influence the absorption probability for the XUV radiation, since the IR intensity is too low to ionize or excite He from the ground state. For large negative delays, when the APT precedes the IR pulse, effects like perturbed free polarization decay occur, and these influence the absorption probability \cite{Chen12}. Hence, the total absorption is not expected to be symmetric around delay-zero. The maximum of the total absorption retrieved by the fitting procedure appears to be 9.4\,fs before the maximum of the 4$\omega$-oscillations. This leads to a completely different calibration of the delay axis. 

Another way to experimentally determine the delay-zero is using a signature based on a nonlinear process of higher order. We use the Gaussian-Wigner transform already presented in the previous section and apply it to the energy-integrated total absorption. By integrating the resulting two-dimensional representation along the frequency axis, we obtain the envelope of the 2$\omega$-oscillations as a function of delay. For the 2$\omega$-oscillations we integrate around 0.76\,PHz with an integration window of 0.6\,PHz. Figure 5(b) presents the result for an IR intensity of $2.6\cdot10^{12}$\,W/cm$^2$. Intuitively, we might expect delay-zero to coincide with the maximum of the nonlinear signature. In contrast, figure 5(b) shows that the 2$\omega$ envelope has an asymmetric shape and splits into two peaks where the maximum is located more than 7 fs before delay-zero. In addition, the asymmetric splitting of the 2$\omega$ envelope strongly depends on the IR intensity as we discuss in the following paragraph. Accordingly, the envelope of the 2$\omega$-oscillations of the energy-integrated total absorption is also not suitable for a precise delay-zero calibration.

\begin{figure}[t]
\begin{center}
 \begin{tabular}{c}
\includegraphics[width=14cm]{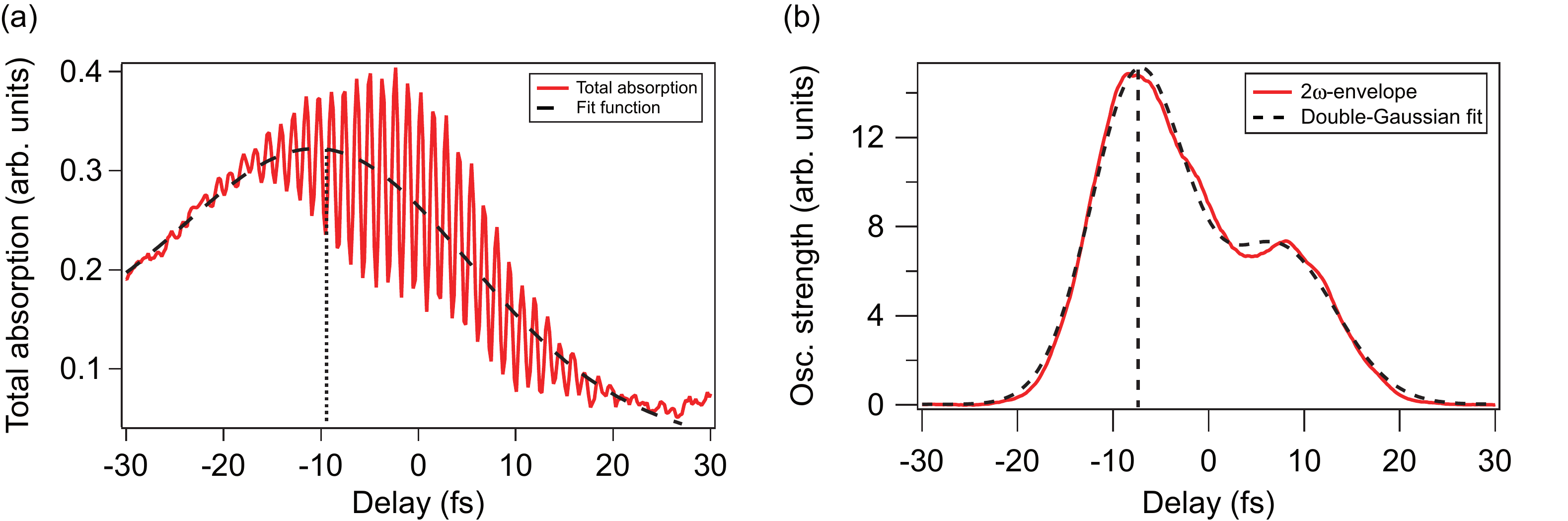}
 \end{tabular}
\end{center}
\caption[example]{\label{fig:Figure_badcalibration} (a) Total absorption as function of APT-IR delay with an IR intensity of $1.3\cdot10^{12}$\,W/cm$^2$. The experimental data are fitted with the sum of a Gaussian and a linear function. The center of the Gaussian in the fit function disagrees with the delay-zero defined with the 4$\omega$-oscillations by 9.4\,fs. (b) Envelope of the 2?-oscillations of the energy-integrated absorption at an IR intensity of $2.6\cdot10^{12}$\,W/cm$^2$. The envelope shows an asymmetric shape and the maximum is located more than 7\,fs before the delay-zero.} 
\end{figure}

A further idea for a delay-zero calibration is the application of the time-frequency analysis to individual HHs instead of investigating the total energy-integrated absorption. Figure 6 presents the envelope of the 2$\omega$-oscillations for HH 13, 15 and 17 for different IR intensities. The envelopes of all three harmonics exhibit a strong dependence on the IR intensity regarding the oscillation strength and the shape. For the lowest intensity shown here, $1.3\cdot10^{12}$\,W/cm$^2$, the envelope is symmetrically centered on the delay-zero defined through the maximum of the 4$\omega$-oscillations of HH 13, as in the case of the 4$\omega$-oscillations shown in figure 7. With increasing IR intensity the symmetric envelope starts to split up into two peaks. Furthermore, the amplitude of the envelope decreases. It is also important to note that the splitting of the envelope is asymmetric, in the sense that one of the two peaks is dominant (the description of the asymmetric shape of the envelope is beyond the scope of the work presented here). For HHs 13 and 17 we observe that the peak at negative delays is on average higher, whereas the dominant peak for HH 15 is located at positive delays. Both of these observations are reproduced in the calculations. This complex behavior of the envelope shows that the maximum of the 2$\omega$-oscillations is not suitable for a correct delay-zero calibration. Indeed, the maximum peak can be shifted up from the real delay-zero by as much as 20\,fs in the intensity range covered with our measurements.

\begin{figure}[t]
\begin{center}
 \begin{tabular}{c}
\includegraphics[width=14cm]{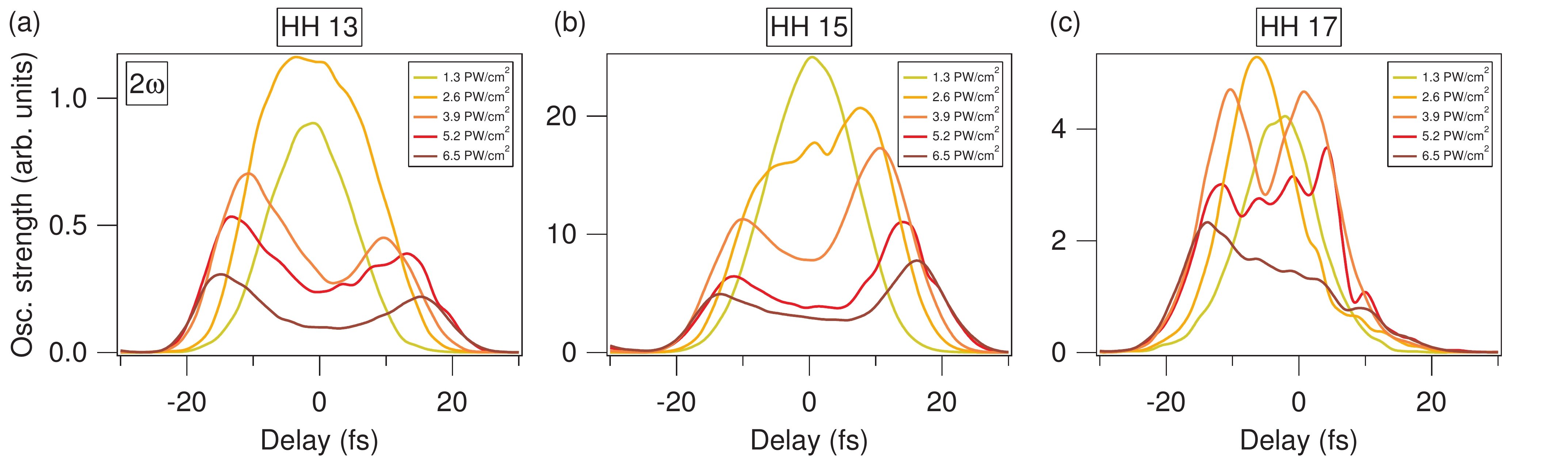}
 \end{tabular}
\end{center}
\caption[example]{\label{fig:Figure_2wenvelope} Intensity dependence of the 2$\omega$-oscillationsÕ envelope for HH 13 (a), 15 (b) and 17 (c) obtained by integrating the Gaussian-Wigner transform shown in figures 4(a)-(c) in the frequency domain around 0.76\,PHz. The width of the integration window is 0.6\,PHz.}
\end{figure}

The 4$\omega$-oscillations in the transient absorption signal result from the highly nonlinear coupling of two HHs via four IR photons. For the envelope of the 4$\omega$-oscillations we integrate the Gaussian-Wigner transform in the frequency domain, as we did for the 2$\omega$-oscillations envelope. The integration window again has a width of 0.6\,PHz but is centered on a frequency of 1.52\,PHz. Figure 7 presents the envelope of the 4$\omega$-oscillations for HH 13 and 17 at two different IR intensities, showing both theoretical ((a) and (b)) and experimental ((c) and (d)) results. In the calculations, delay-zero is known exactly by definition. The theoretical results show that the envelope of the 4$\omega$-oscillations is centered very accurately at delay-zero and possesses a symmetric shape as a function of delay. In the calculations, we observe that this behavior is independent of the IR intensity over the whole range of moderate intensities studied. The experimental results in figure 7(c) and (d) also exhibit a symmetric envelope for the 4$\omega$-oscillations which is not affected by the IR intensity in our IR-intensity range. The symmetric shape enables us to fit the envelope with a Gaussian. The fit function provides, besides the peak value, which we use to define the strength of the oscillations, also a peak position. The peak position is of special interest for us to experimentally define the delay-zero. The excellent agreement between the symmetric envelopes of the 4$\omega$-oscillations of the experimental and the theoretical results suggests that the experimentally measured position of the maximum of the 4$\omega$-oscillations in HH 13 and 17 is an appropriate determination of exact delay-zero. 

\begin{figure}[t]
\begin{center}
 \begin{tabular}{c}
\includegraphics[width=14cm]{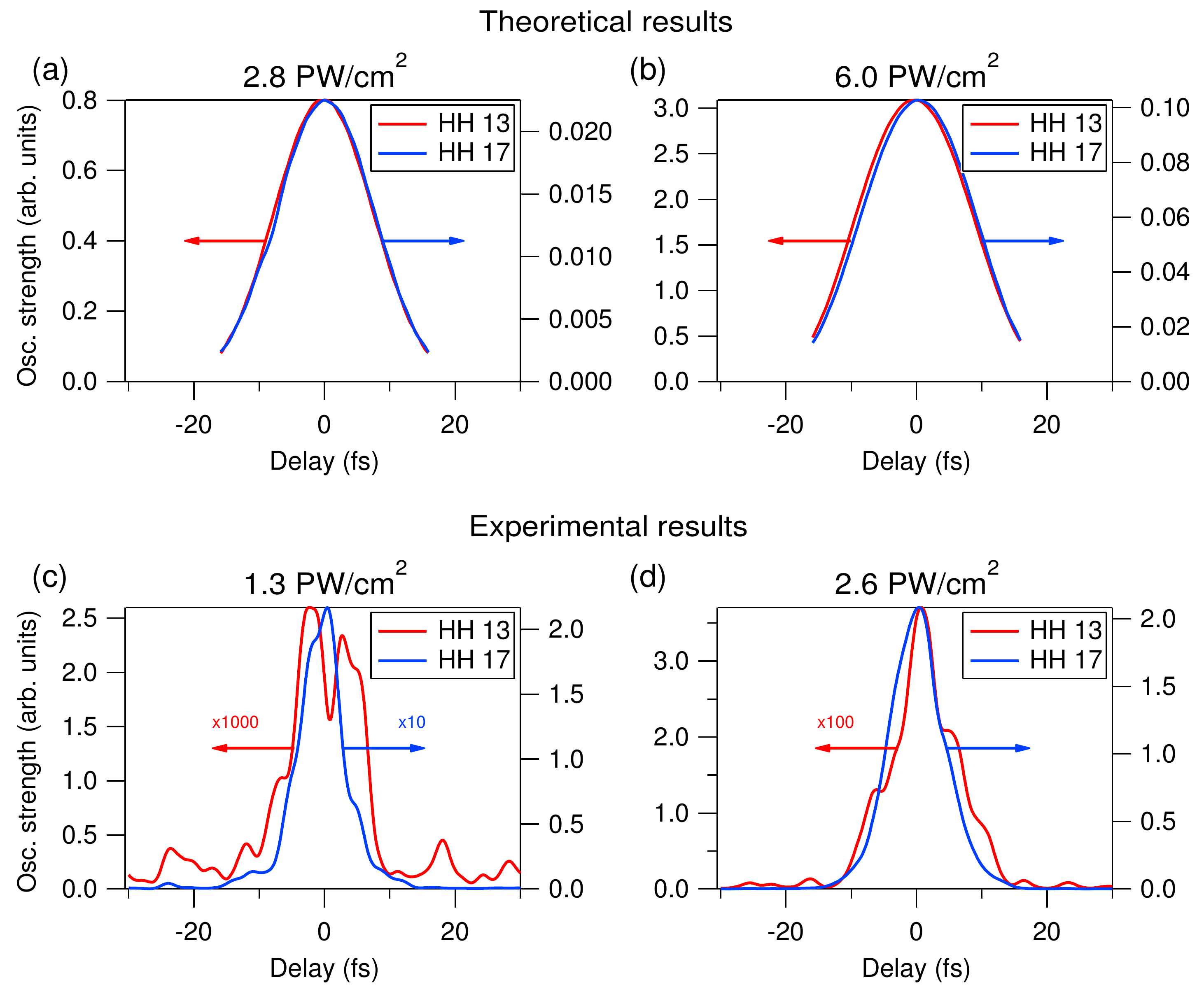}
 \end{tabular}
\end{center}
\caption[example]{\label{fig:Figure_4wenvelope} (a) Calculated temporal envelope of the 4$\omega$-oscillations of HH 13 (red curve) and 17 (blue curve) at IR intensity of $2.8\cdot10^{12}$\,W/cm$^2$ and (b) of $6.0\cdot10^{12}$\,W/cm$^2$. The envelope of the 4$\omega$-oscillations is symmetrically centered around delay-zero and the position of its maximum is for both HHs independent of the IR intensity. (c) and (d) show the envelope of the corresponding experimental results for an IR intensity of $1.3\cdot10^{12}$\,W/cm$^2$ and $2.6\cdot10^{12}$\,W/cm$^2$.}
\end{figure}

\section{IR-intensity dependence}

As discussed by Chen and co-workers, we expect an IR-intensity dependence for the magnitude of the oscillations of the 2$\omega$- and 4$\omega$-oscillations [11]. The motorized iris in the IR beam path enables us to vary the IR intensity in the He target between $1.3\cdot10^{12}$\,W/cm$^2$ and $8.6\cdot10^{12}$\,W/cm$^2$. These intensities are insufficient to induce any transition out of the He ground state to bound excited states or into the continuum. In order to quantify the magnitude of the oscillations, we extract the information on their strength by fitting the temporal envelope of the 2$\omega$- and 4$\omega$-oscillations. The symmetric shape of the 4$\omega$-oscillation strength as a function of APT-IR delay is, as already mentioned in section 3, fitted with a Gaussian. Conversely, as we showed in section 3 and in figure 6, the 2$\omega$-oscillations have an asymmetric envelope, which strongly depends on the IR intensity. Hence, we fit this envelope with a sum of two Gaussians. Figure 8 shows the resulting strengths of the 2$\omega$- and 4$\omega$-oscillations normalized to the value we obtain at the lowest IR intensity of $1.3\cdot10^{12}$\,W/cm$^2$. The 2$\omega$-component of HH 13 and 17 shows a local maximum around $\sim4.0\cdot10^{12}$\,W/cm$^2$. If we increase the IR intensity even further the oscillation strength declines. For HH 15 we observe a monotonic decrease of the 2$\omega$-oscillation strength over the full scanned intensity range.

The intensity scan for the 4$\omega$-oscillations of HH 13 and 17 (figure 8(b)) shows two distinguishable intensity ranges. For intensities up to $\sim4.0\cdot10^{12}$\,W/cm$^2$ the normalized oscillation strength rises monotonically. If we further increase the intensity we enter a different regime, where the normalized oscillation strength is decreasing with increasing intensity. 

In the calculations, we do not observe this behavior when using an IR wavelength of 795\,nm. Rather, we find that the 4$\omega$-oscillation strengths of HH 13 and HH 17 increase monotonically with intensity. However, in the single atom calculations we have explored changing the IR wavelength (which means that the harmonic wavelengths, which are locked to the IR wavelength, also change) and then studying the intensity dependence of the HH13 and HH17 total absorption, and their respective 2$\omega$- and 4$\omega$-components. Figure 9 shows an example of this 2D exploration of parameter space for the 4$\omega$-component of HH 17 (a), and for the total absorption of HH 13 (b) and HH 15 (c). Figure 9(a) shows that at the shortest IR wavelengths there are two dominant structures in the 4$\omega$-component of HH 17. These can be identified as the resonant enhancement of the absorption due to the 2p state (resonant with HH 13, as shown in figure 9(b)) and the Stark-shifted 3p state (resonant with HH 15, and thus a two-IR-photon intermediate resonance for the HH 13-HH 17 coupling). Where the two resonances meet, and the excitation dynamics are therefore very complex, there is a wavelength regime in which the 4$\omega$-oscillation strength decreases with intensity. This wavelength is slightly shorter (775\,nm) than that used in the experiment (789\,nm). One can speculate that the experimental APT harmonics could possibly have been blue shifted in the generating Xe jet due to plasma self-phase modulation, or that theory possibly does not accurately predict the Stark shift of the 3p state. 

\begin{figure}[t]
\begin{center}
 \begin{tabular}{c}
\includegraphics[width=14cm]{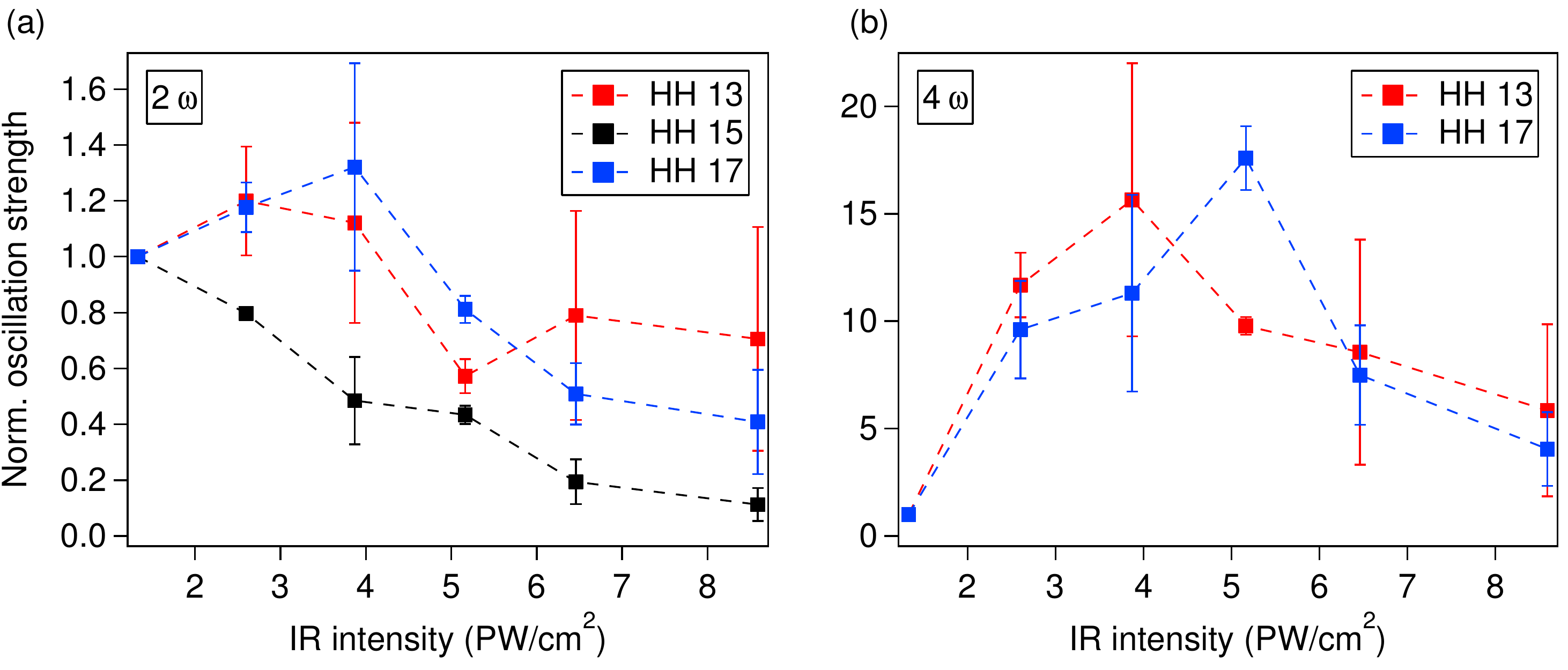}
 \end{tabular}
\end{center}
\caption[example]{\label{fig:Figure_IRintensity} (a) Intensity dependence of the normalized oscillation strength of the 2$\omega$-oscillations for HHs 13 to 17 and (b) 4$\omega$-oscillations for HHs 13 and 17. The oscillations strength is normalized to value we obtain at the lowest IR intensity of $1.3\cdot10^{12}$\,W/cm$^2$.}
\end{figure}

\begin{figure}[t]
\begin{center}
 \begin{tabular}{c}
\includegraphics[width=14cm]{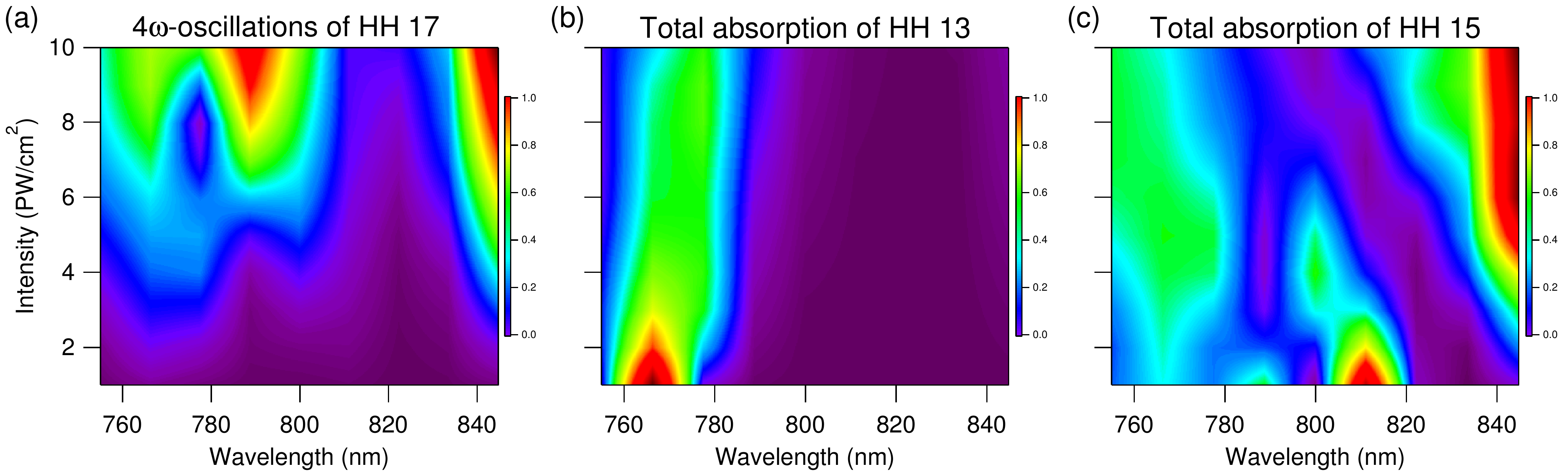}
 \end{tabular}
\end{center}
\caption[example]{\label{fig:Figure_calculation} Single atom calculations of absorption yields as a function of laser wavelength and peak intensity. (a)  4$\omega$-oscillations amplitude in the absorption of HH 17 (note that a similar plot for HH 13 would look almost identical), (b) total absorption of HH 13, (c) total absorption of HH 15. }
\end{figure}

\section{Photoabsorption and photoionization probabilities}

As described in reference \cite{Locher14}, our setup also allows for the detection of charged particles. For this purpose we remove the pulsed gas target from the interaction region and insert a needle target. This target provides a continuous gas flow with a low gas load so that we can operate a time-of-flight (TOF) spectrometer with a micro-channel plate detector. 

Figure 10 shows the He\textsuperscript{+} ion yield as a function of APT-IR delay (black curve). In order to avoid the situation in which the dominant contribution in the ion yield comes from the harmonics above the first ionization potential, we adjusted the XUV spectrum of the APT. We changed the intensity of the generating field and obtained a spectrum dominated by HHs 13 and 15. In this way the relative IR-induced changes in the photoionization become stronger and the analysis of the oscillating signal is more robust. As can be seen in figure 10(a), the ion yield has a maximum when IR and APT overlap around delay-zero, which mostly follows the maximum of the integrated absorption. In this region direct ionization by the harmonics above the ionization threshold and multiphoton ionization by harmonics below the threshold in combination with IR photons contribute to the ion yield. For large negative delays, when the APT precedes the IR pulse, we detect a higher ion yield than for large positive delays, where the IR field arrives first. In the case of large negative delays, the IR can ionize states which were populated by the preceding APT. This is in agreement with earlier results \cite{Ranitovic11}. Additionally, we show in figure 10 the total absorption probability of the XUV radiation integrated in energy from 19\,eV to 35\,eV (red curve). Both in the ion yield and in the total absorption probability we observe strong 2$\omega$-oscillations. In order to verify the appearance of 4$\omega$-oscillations we perform the delay-frequency analysis with the help of the Gaussian-Wigner transform as described earlier. This analysis shows that both the ion and the optical signal exhibit 2$\omega$- and 4$\omega$-oscillations. We again integrate the delay-frequency representation in the frequency domain to obtain the envelope of the oscillations as a function of delay. Figure 10(b) shows the result for the 4$\omega$-oscillations of the ion yield and the total absorption probability as a function of APT-IR delay. The modulations in the envelope of the ion signal are a numerical artifact of the Gaussian-Wigner transform. We fit both envelopes with a Gaussian, shown as dashed lines, and define delay-zero with the center of the Gaussian fit for the absorption probability. As shown in figure 10(b), determining delay-zero from the ion signal in the same fashion (black dotted line) yields good agreement with our definition of the delay-zero extracted from the optical response. A deviation $\Delta\tau$ of less than 1\,fs is measured. 

\begin{figure}[t]
\begin{center}
 \begin{tabular}{c}
\includegraphics[width=14cm]{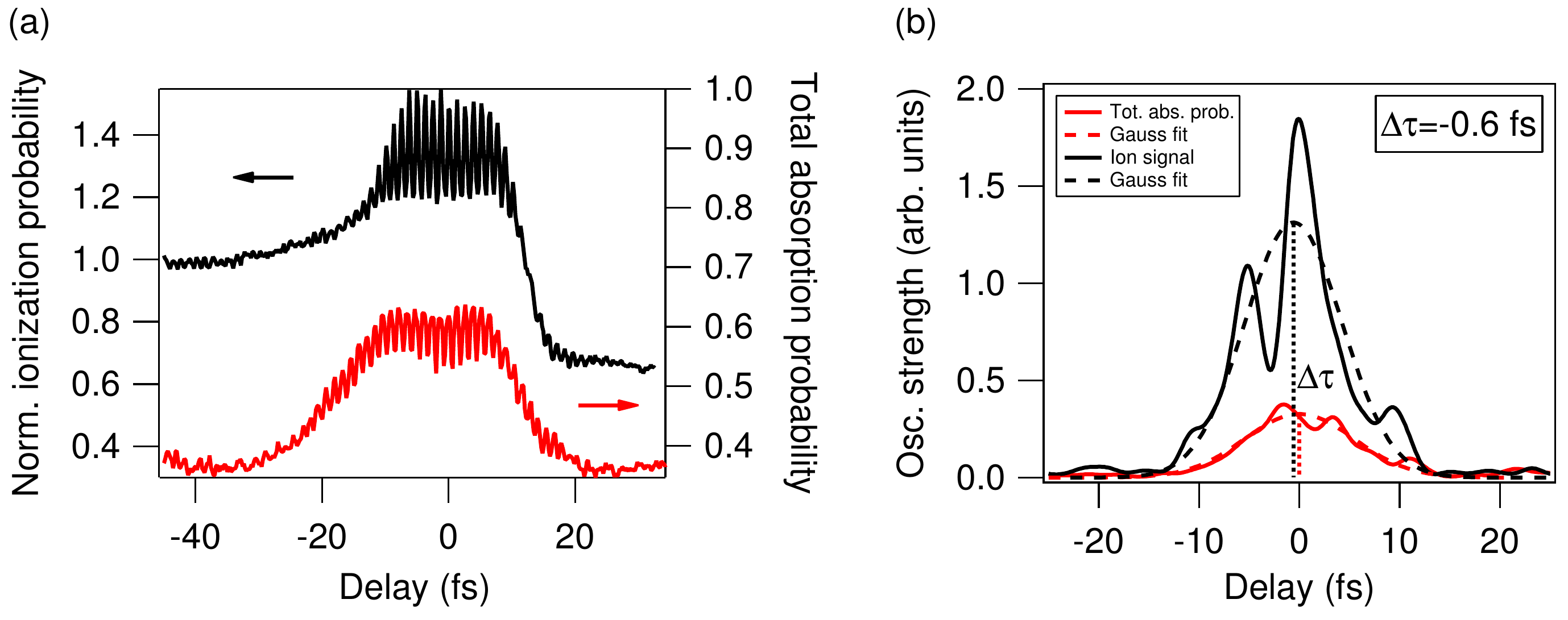}
 \end{tabular}
\end{center}
\caption[example]{\label{fig:Figure_ions} (a) He\textsuperscript{+} ion yield (black curve) and total absorption probability (red curve) as a function of APT-IR delay. (b) Envelope of 4$\omega$-oscillations of the ion yield (black curve) and total absorption probability (red curve). The dashed lines represent the Gaussian fit functions. The position of the maximum of the fit function is indicated by the vertical dotted line. The maximum of the fit function for the ion signal exhibits a deviation $\Delta\tau$ of the delay-zero defined by the maximum of the 4$\omega$-oscillations of the absorption probability of less than 1 fs.}
\end{figure}

\section{Conclusions}

In conclusion, we have measured and characterized sub-cycle oscillations in an attosecond transient absorption experiment in He by using an APT and a moderately strong femtosecond IR pulse. In addition to the earlier observed 2$\omega$-oscillations, we observed 4$\omega$-oscillations, a periodicity which is not included in the initial interacting fields. We show that the 4$\omega$-oscillations in the transient absorption signal can be used to determine delay-zero with an accuracy which is better than other experimental methods, including the total absorption and 2$\omega$-oscillations. A systematic investigation of the IR dependency of the 4$\omega$-oscillations reveals the influence of resonances on the oscillation strength. These experimental results are in excellent agreement with TDSE calculations. Additionally, we compare the total absorption probability of the APT with the He\textsuperscript{+} ion yield. The ion yield exhibits, like the total absorption probability, 2$\omega$- and 4$\omega$-oscillations and the position of the maximum of the 4$\omega$-oscillations is in agreement with the transient absorption measurement. Consequently, the calibration of the delay-zero based on the 4$\omega$ signature is not restricted to attosecond transient absorption spectroscopy but may also be very helpful for experiments based on the detection of charged particles. 

\ack
This research was supported by the ETH Zurich Postdoctoral Fellowship Program and the NCCR MUST, funded by the Swiss National Science Foundation, and by the Office of Science, Office of Basic Energy Sciences, Geosciences, and Biosciences Division of the US Department of Energy under Contract No. DE-FG02-13ER16403. High-performance computing resources were provided by the Louisiana Optical Network Initiative (LONI). 

\section*{References}


\begin{thebibliography}{10}

\bibitem{Gallmann13}
Gallmann L \textit{et al} 2013 {\it Molecular Physics \/}{\bf 111}
2243

\bibitem{Goulielmakis10}
Goulielmakis E \textit{et al} 2010 {\it Nature \/}{\bf 466}
739

\bibitem{Wang10}
Wang H \textit{et al} 2010 {\it Phys. Rev. Lett.\/}{\bf 105}
143002

\bibitem{Holler11}
Holler M, Schapper F, Gallmann L and Keller U 2011 {\it Phys. Rev. Lett. \/}{\bf 106}
123601

\bibitem{Chini12}
Chini M, \textit{et al} 2012 {\it Phys. Rev. Lett.\/} {\bf 109}
073601

\bibitem{Chen12}
Chen S \textit{et al} 2012 {\it Phys. Rev. A\/} {\bf 86}
063408

\bibitem{Chini13}
Chini M \textit{et al} 2013 {\it Sci. Rep.\/} {\bf 3}
1105

\bibitem{Pfeiffer13}
Pfeiffer A N \textit{et al} 2013 {\it Phys. Rev. A\/} {\bf 88}
051402(R)

\bibitem{Lucchini13}
Lucchini M \textit{et al} 2013 {\it New Journal of Physics\/} {\bf 15}
103010

\bibitem{Herrmann13}
Herrmann J \textit{et al} 2013 {\it Phys. Rev. A\/} {\bf 88}
043843

\bibitem{Chen12_2}
Chen S, Schafer K J and Gaarde M B 2012 {\it Optics Letters\/} {\bf 37}
2211

\bibitem{Chen13}
Chen S, Wu M, Gaarde M B, Schafer K J 2013 {\it Phys. Rev. A\/} {\bf 88}
033409

\bibitem{Wang13}
Wang X \textit{et al} 2013 {\it Phys. Rev. A\/} {\bf 87}
063413

\bibitem{Johnsson07}
Johnsson P \textit{et al} 2007 {\it Phys. Rev. Lett.\/} {\bf 99}
233001

\bibitem{Locher14}
Locher R \textit{et al} 2014 {\it Review of Scientific Instruments\/} {\bf 85}
013113

\bibitem{Lopez05}
L\'opez-Martens, R \textit{et al} 2005 {\it Phys. Rev. Lett.\/} {\bf 94}
033001

\bibitem{Wiese09}
Wiese W L and Fuhr J R 2009 {\it J. Phys. Chem. Ref. Data\/} {\bf 38}
565

\bibitem{Wigner32}
Wigner E P 1932 {\it Phys. Rev.\/} {\bf 40}
749

\bibitem{Gaarde11}
Gaarde M B, Buth C, Tate J L and Schafer K J 2011 {\it Phys. Rev. A\/} {\bf 83}
013419

\bibitem{Ranitovic11}
Ranitovic P \textit{et al} 2011 {\it New Journal of Physics\/} {\bf 12}
013008

\end{thebibliography}
\end{document}